\documentclass{elsart1p}
\usepackage[latin1]{inputenc}

\usepackage{graphicx,amsmath,amssymb, amsfonts}
\begin{document}

\begin{frontmatter}

\title{Vlasov simulations of collisionless magnetic reconnection without
background density}
       
\author{H. Schmitz and R. Grauer}

\address{Theoretische Physik I, Ruhr-Universit\"at Bochum, 44780 Bochum,
Germany}

\begin{abstract} A standard starting point for the simulation of
collisionless reconnection is the Harris equilibrium which is made up of a
current sheet that separates two regions of opposing magnetic field.
Magnetohydrodynamic simulations of collisionless reconnection usually
include a homogeneous background density for reasons of numerical
stability. While, in some cases, this is a realistic assumption, the
background density may introduce new effects both due to the more involved
structure of the distribution function or due to the fact that the Alfv\`en
speed remains finite far away from the current sheet. We present a fully
kinetic Vlasov simulation of the perturbed Harris equilibrium using a
Vlasov code. Parameters are chosen to match the Geospace Environment
Modeling (GEM) Magnetic Reconnection Challenge but excluding the background
density. This allows to compare with earlier simulations [Schmitz,
Grauer, Phys.\ Plasmas 13 (2006) 092309] which include the background
density. It is found that the absence of a background density causes the
reconnection rate to be higher. On the other hand, the time until the onset
of reconnection is hardly affected. Again the off diagonal elements of the
pressure tensor are found to be important on the X--line but with modified
importance for the individual terms. \end{abstract}

\begin{keyword}

02.70.-c 
52.25.Dg 
52.65.Ff 
52.25.Xz 

\end{keyword}

\end{frontmatter}

\section{Introduction}

Magnetic reconnection is widely believed to be the most important process
for converting magnetic energy into kinetic energy of the plasma. It is
relevant for both space and laboratory plasmas and thought to be the main
player for energy release in solar flares, coronal mass ejections and
substorms in the earth's magnetosphere.

The models used to investigate collisionless magnetic reconnection usually
start from the Harris--equilibrium. In this equilibrium two regions of
oppositely oriented magnetic field lines are separated by a current sheet.
A small perturbation of this equilibrium leads to a localised narrowing of
the current sheet until the field lines reconnect on the X--line. In the
ideal MHD model, the frozen field condition prohibits such a reconnection.
Therefore the reconnection process depends on a nonideal mechanism that
breaks the frozen field condition. In the last years it has become apparent
that the Hall--MHD framework provides a minimum model in understanding the
fast reconnection (see eg.
\cite{Birn:2001a,Shay:01,Lottermoser-Scholer:1997}). The Hall--Term itself,
however, cannot provide this mechanism, since the field lines
are still frozen in the electron flow. Due to symmetry constraints, only
the electron inertia and the off--diagonal terms of the pressure tensor
can provide a nonideal mechanism on the X--line itself (see e.g Refs.\
\cite{Vasyliunas:1975,Dungey:1988,Hesse:1993,Kuznetsova:1998}). 

In this study, we use a Vlasov code to investigate collisionless
reconnection without background density which is relevant, for example, for
magnetotail reconnection. Apart from the background density, all other
parameters are identical to the GEM setup to allow comparison. In the next
section we briefly present the methods used in our investigation. The setup
including the initial conditions and the boundary conditions is described
in section \ref{SecSetup}. In section \ref{SecResults} we discuss the
results. Separate subsections are dedicated to the discussion of the
contributions of the terms in Ohm's law, especially the off--diagonal
components of the pressure tensor, and to the detailed discussion of the
electron distribution function. Section \ref{SecSummary} will give a
summary and present conclusions.

\section{Methods\label{SecMethods}}

The kinetic description starts from the distribution functions $f_k({\bf
x}, {\bf v},t)$ of species $k$, where $k=i,e$ denotes ions or electrons. 
The time development of the distribution function is described by the Vlasov
equation
\begin{displaymath}
\frac{\partial f_k}{\partial t} 
        + {\bf v}\cdot \nabla f_k
        + \frac{q_k}{m_k} \left( 
                {\bf E} + {\bf v} \times {\bf B}
        \right) \cdot \nabla_{\bf v} f_k 
= 0.\label{VlasovOrig}
\end{displaymath}
Here $q_k$ and $m_k$ are the charge and the mass of the particles of species
$k$. The Vlasov equation describes the incompressible flow of the species
phase space densities under the influence of the electromagnetic fields.

The electromagnetic fields are solved using the Darwin approximation (see,
for example Refs.\ \cite{BIR85,SCHM06a}). The elimination of the vacuum
modes in the Darwin approximation allows larger time-steps in the
simulation since only the slower non--relativistic waves have to be
resolved.

To close the set of equations of the Vlasov--Darwin system the charge
density $\rho$ and the current density $\mathbf{j}$ have to be calculated
from the distribution function,
\begin{equation}
\rho = \sum_k q_k \int f_k(\mathbf{x}, \mathbf{v}) d^3v \;\; ,
  \qquad\qquad
\mathbf{j} = \sum_k q_k \int \mathbf{v} f_k(\mathbf{x}, \mathbf{v}) d^3v \;\; .
\end{equation}

We use a $2\frac{1}{2}$--dimensional Vlasov--code described in Refs.\
\cite{SCHM06a,SCHM06c}. The term $2\frac{1}{2}$--dimensio\-nal  means,
we restrict the simulations to 2 dimensions in space but include all three
velocity dimensions. The integration scheme is based on a flux conservative
and positive scheme \cite{Filbet:2001} which obeys the maximum principle and
suffers from relatively little numerical diffusion. 

\section{Setup\label{SecSetup}}
The reconnection setup is identical to the parameters of the GEM magnetic
reconnection challenge \cite{Birn:2001a}. The initial conditions are based on
the Harris sheet equilibrium \cite{HARS62} in the $x$,$y$--plane
\begin{equation}
\mathbf{B}(y) = B_0 \tanh\left( \frac{y}{\lambda} \right)\mathbf{\hat{x}}.
\end{equation}
The particles have a shifted Maxwellian distribution 
with constant electron and ion temperatures $T_{i,e}$  and
constant electron and ion drift velocities $V_{0 i,e}$. The
density distribution is then given by
$n_0(y) = n_0 \text{ sech}^2\left( {y}/{\lambda} \right)$.

The total system size is $L_x=25.6\lambda_i$ by 
$L_y=12.8\lambda_i$, where $\lambda_i$ is the ion
inertial length. Because of the symmetry constraints we simulate only one
quarter of the total system size: $0\le x \le L_x/2$ and $0\le y \le
L_y/2$. The sheet half thickness is chosen to be $\lambda=0.5\lambda_i$.
The temperature ratio is $T_e/T_i = 0.2$ and a reduced mass ratio of
$m_i/m_e = 25$ is used. The simulation is
performed on  $256\times128$ grid points in space for the quarter
simulation box. This corresponds to a resolution of $512\times256$. This
implies a grid spacing of $\Delta x = \Delta y = 0.05 \lambda_i$. The
resolution in the velocity space was chosen to be $30\times30\times30$ grid
points. The simulation was performed on a 32 processor Opteron cluster and
took approximately 150 hours to complete.

\begin{figure}[t]
\begin{center}
\includegraphics[width=8.3cm,height=5.1cm]{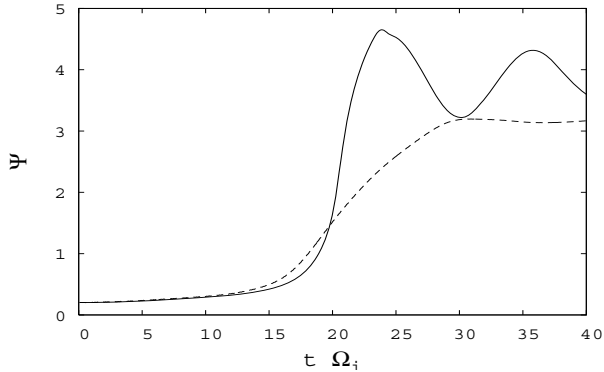}
\end{center}
\caption{Time evolution of the reconnected magnetic flux $\Psi$ throughout
the simulation run without background density (solid line) and with
background density (dashed line). \label{FigRecFlux}}
\end{figure}

An initial perturbation 
\begin{equation}
\psi(x,y) = \psi_0\cos(2\pi x/L_x)\cos(\pi y/L_y)
\end{equation}
is added to the magnetic vector potential component $A_z$. To place the
system directly into the nonlinear regime, the magnitude of
the perturbation is chosen to be $\psi_0 = 0.1 B_0/\lambda_i$. We are not
interested in the linear growth of the instability, but rather in the
nonlinear reconnection phase that follows.

\section{Simulation Results\label{SecResults}}

Figure \ref{FigRecFlux}  shows the reconnected flux $\Psi = \int_O^X B_y
dx$  against time throughout the simulation. For comparison, the flux for
the simulation with a background density $n_{\infty} = 0.2$ (see Ref
\cite{SCHM06c}) is also shown. One can observe that the onset of the
reconnection takes place roughly at the same time for both cases, although
it is slightly delayed for the $n_{\infty} = 0$ case. In the further
development the $n_{\infty} = 0$ case shows a reconnection which is much
faster than with a background density. This can be understood by the
increase of the Alfv\`en speed as the density is reduced. In addition, one
can observe oscillations of the reconnected flux for the $n_{\infty}=0$
case after the fast reconnection phase. The average value during the
first oscillations is higher than the highest value of the $n_{\infty} \ne
0$ case. When reconnection stops, the plasma is almost completely trapped
in the magnetic island. During reconnection the ions are accelerated
towards the centre of the island (O--point). After reconnection stops the
ions overshoot due to their inertia and the island starts oscillating. This
oscillation is visible in the reconnected flux for times $t\Omega_i \ge
25$.

\begin{figure}[t]
\begin{center}
Electron out of plane current $j_{e,z}$\\
\includegraphics[width=8.7cm]{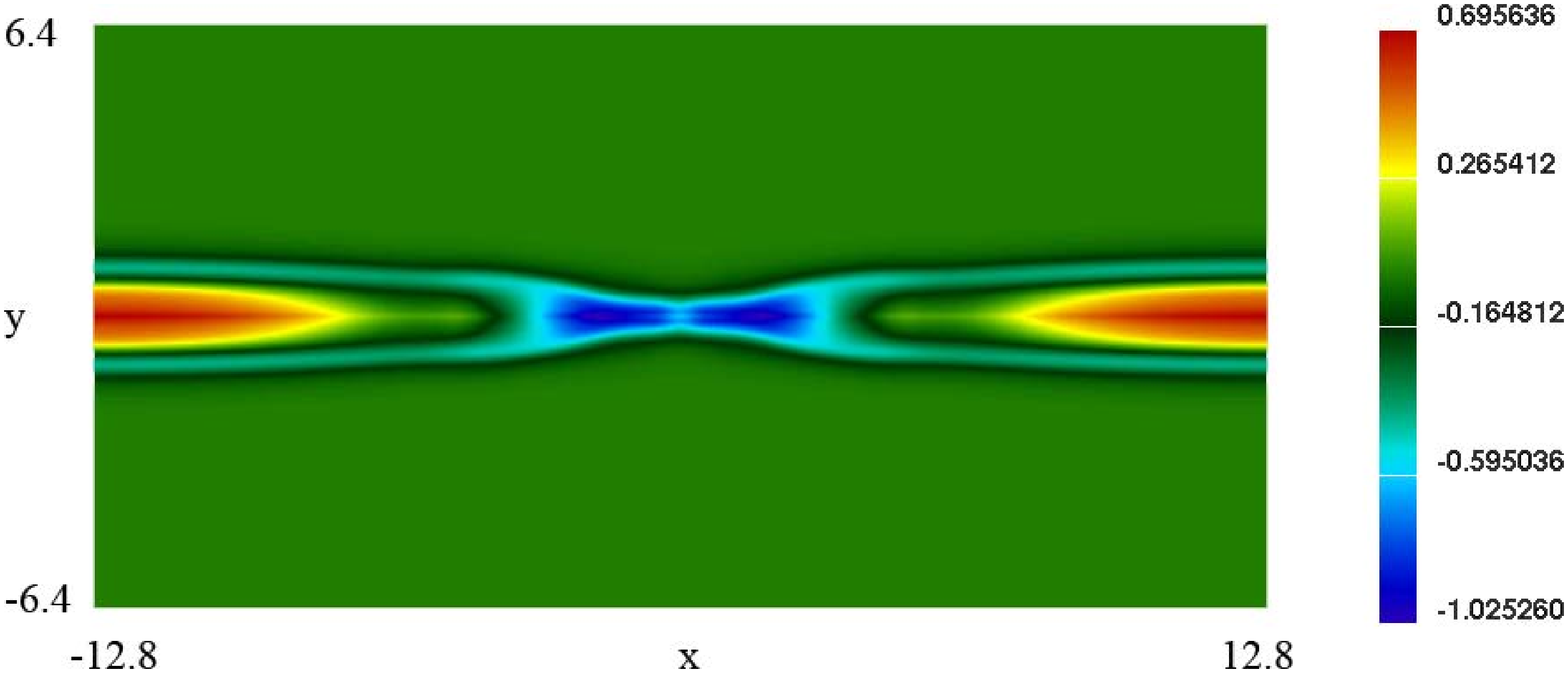}\\[4mm]
Ion out of plane current $j_{i,z}$\\
\includegraphics[width=8.7cm]{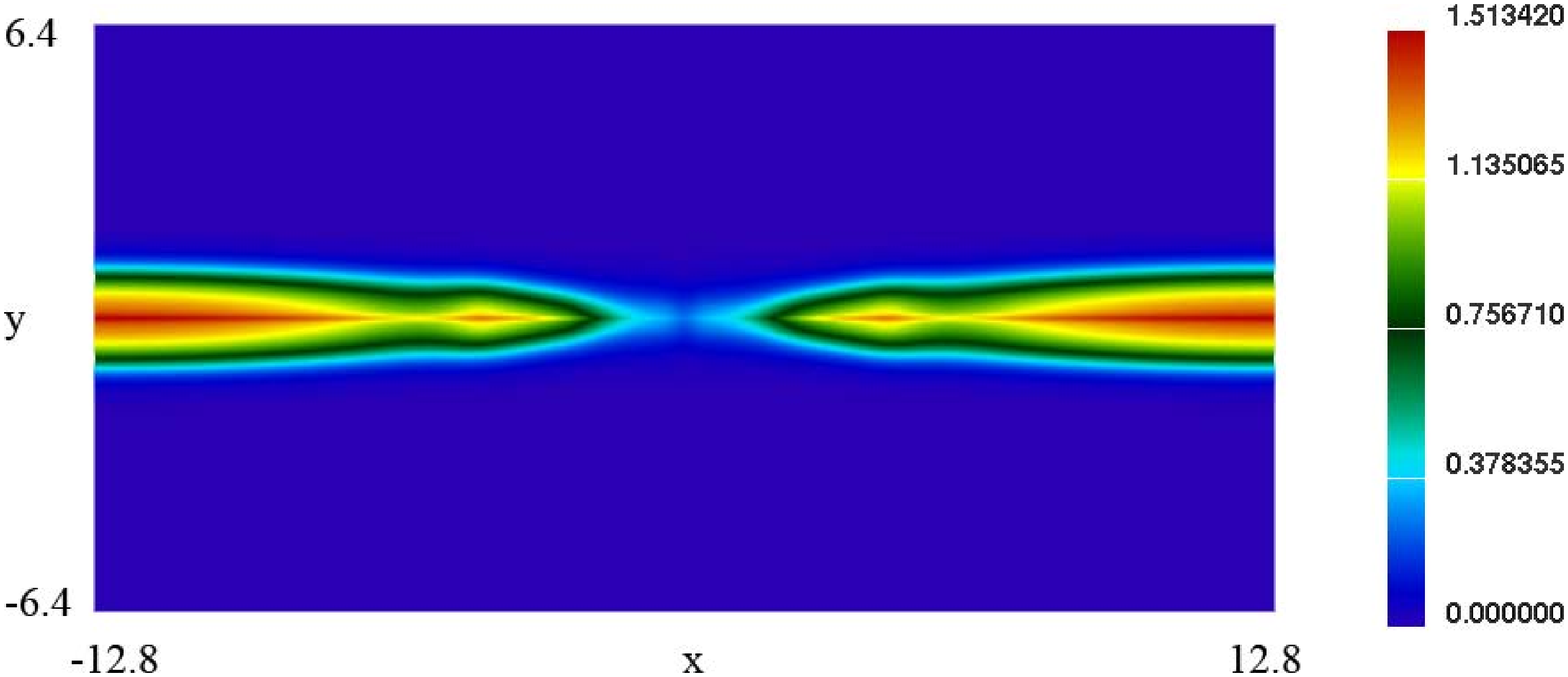}\\[4mm]
\end{center}
\caption{The electron out of plane current $j_{e,z}$ (upper panel) 
and the ion out of plane current $j_{i,z}$ (lower panel) at time 
$\Omega_i t = 18.9$\label{FigOutPlane}}
\end{figure}

Figure \ref{FigOutPlane} shows the out of plane electron and ion current
densities. The figures are plotted at time $\Omega_i t = 18.9$, when the
reconnected flux reaches a value of $\Psi=1$. The features here are similar
to the case with background density \cite{SCHM06c} but some differences can
be seen. The electron current density is decreased on the X--line compared
to the adjacent regions in the diffusion region. This dip in the electron
current is largely due to the lack of current carrying electrons. Both
electron and ion density are almost zero on the X--line. Also the electron
current along the separatrix is not as pronounced as in the $n_{\infty} =
0.2$ case. Here the lack of current carrying electrons at the edges of the
magnetic island is responsible. Finally the lower electron density outside
the island is also the reason for a lower value of the quadrupolar out of
plane magnetic field $B_z$ (not shown). The maximum value of $B_z$ at time
$\Omega_i t = 18.9$ in this simulation is about 0.064 while for the
$n_{\infty} = 0.2$ case it was about 0.165. 

The ion current density $j_{i,z}$ very much follows the particle density
which is roughly equal for both electrons and ions. One can observe a
beginning formation of secondary islands next to the diffusion region. We
attribute this to the high growth rate of the instability for the low
density values. At later times the secondary island coalesces with the main
island.

\begin{figure}
\begin{center}
$P_{xz} / ne$\\
\includegraphics[width=8.7cm]{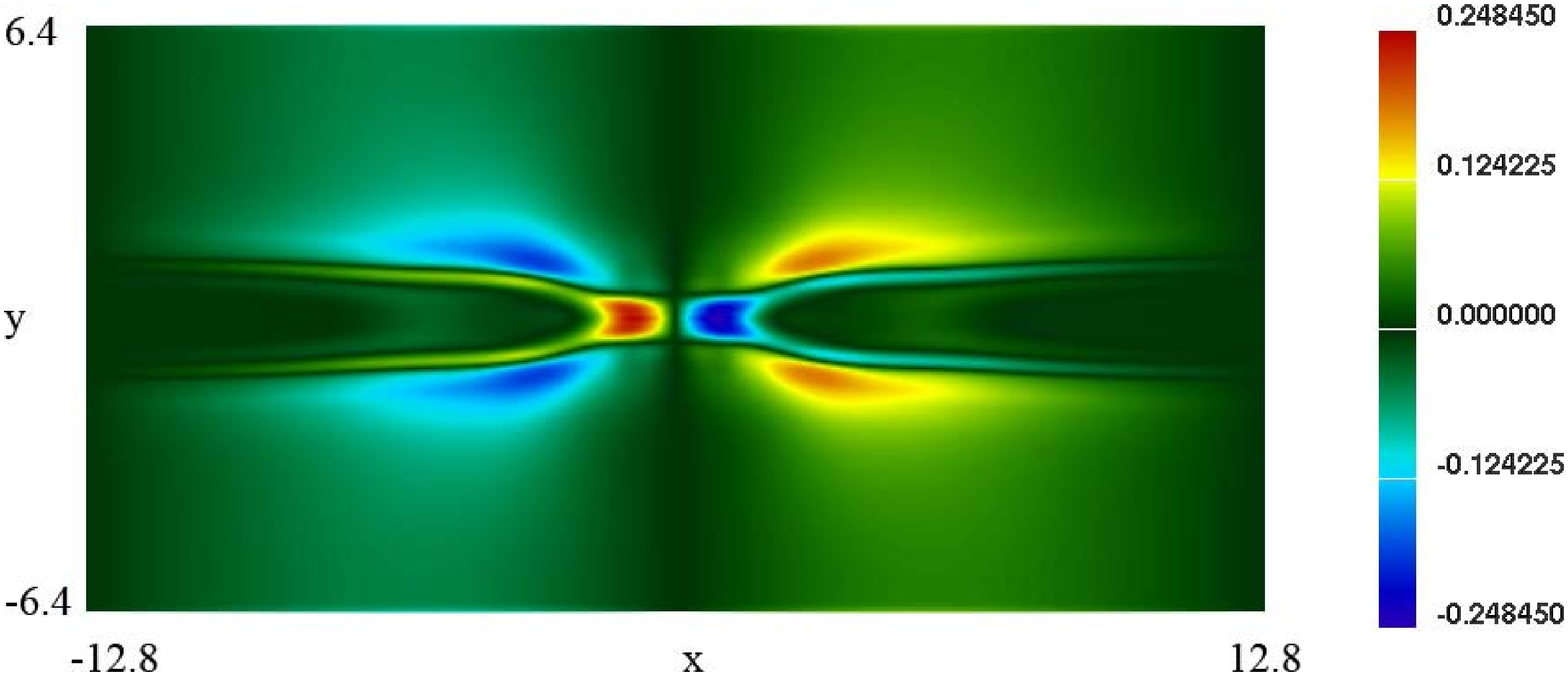}\\[4mm]
$P_{yz} / ne$\\
\includegraphics[width=8.7cm]{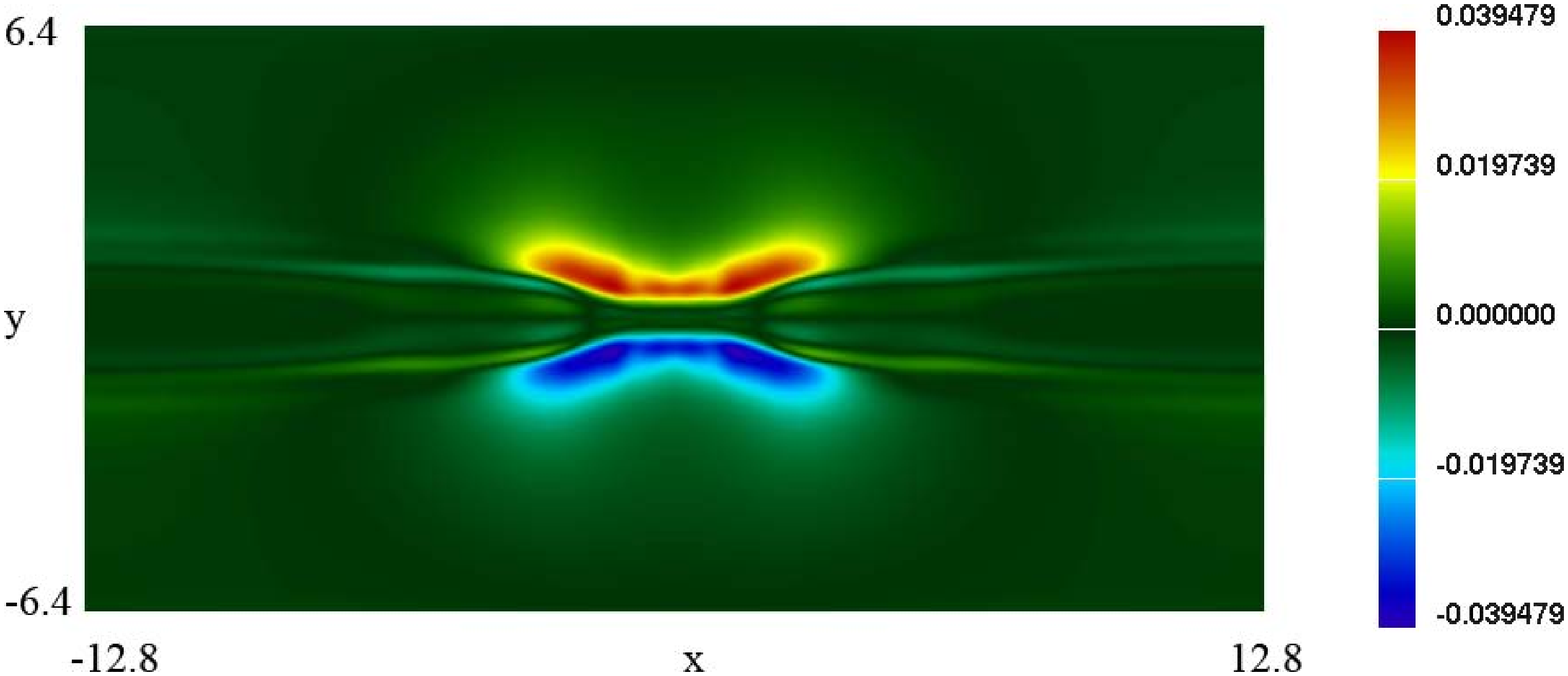}
\end{center}
\caption{The off--diagonal components $P_{xz}$ and $P_{yz}$ of the pressure 
tensor divided by the density $n$ at time $\Omega_i t = 18.9$ 
\label{FigPressureNDiag}}
\end{figure}

In Figure \ref{FigPressureNDiag} we plot the off diagonal components
$P_{xz}$ and $P_{yz}$  of the electron pressure tensor divided by the
density. These two components are the main origin of the inductive electric
field $E_z$ at the X--line. Due to symmetry conditions the electric field
at the X--line is given by $E_z = (m/ne^2) \partial j_z / \partial t -
(1/ne)(  \partial P_{xz} / \partial x + \partial P_{yz} / \partial y)$. In
\cite{SCHM06c} it was observed, for the case with $n_{\infty} = 0.2$, that
the electron inertia played only a secondary role while the two terms from
the pressure tensor contributed to roughly equal amounts to the electric
field. Here we see that, for the $n_{\infty} = 0$ case, the contribution of
the $P_{xz}$ term dominates over the $P_{yz}$ term. The maxima of
$P_{xz}/ne$ are more than a factor of 6 larger than the maxima of
$P_{yz}/ne$. The bar like structure of $P_{yz}$ is still seen, as in
\cite{SCHM06c}, but the bars are further apart, again reducing the
gradient  $\partial P_{yz} / \partial y$. The $P_{xz}$ originates from the
bunched gyro motion of the accelerated electrons in the outflow magnetic
field. On the other hand in \cite{SCHM06c} the $P_{yz}$ term originated
from the electrons that were accelerated in the inflow region and crossed
the neutral line due to their inertia while being accelerated in the
$z$--direction. Without background density these electrons are missing and
the $P_{yz}$ stays small. As a consequence the source inductive electric
field is made up almost completely by the $\partial P_{xz} / \partial x$
contribution.

\section{Summary and Conclusions\label{SecSummary}}

We have performed a $2\frac{1}{2}$--dimensional Vlasov simulation of
collisionless reconnection without a background density. To allow
comparison, all other parameters were chosen to be equal to the GEM setup
\cite{Birn:2001a}. Some differences were found that could be attributed to
the lack of a background population. The onset of the fast reconnection was
not influenced by the background density but the reconnection rate was
found to be considerably faster when no background population was present.
This increase in the reconnection rate can be attributed to the faster
Alfv\`en velocity as the density decreases. The faster reconnection rate
causes secondary islands to form. However, no full development of secondary
X--lines could be observed as the secondary islands quickly coalesce with
the main island. In previous investigations it became apparent, that on the
X--line only the off diagonal components of the electron pressure tensor
carry a major contribution to the reconnection electric field. When a
background density is present both the $P_{xz}$ and $P_{yz}$ contribute to
almost equal amounts. Due to the lack of inflowing electrons in the
$n_{\infty} = 0$ case, the $P_{yz}$ is greatly reduced and only the
$P_{xz}$ is dominant at the X--line.

\section*{Acknowledgements}

This work was supported by the SFB 591 of the Deutsche
Forschungsgesellschaft. Access to the JUMP multiprocessor computer at
Forschungszentrum J\"ulich was made available through project HBO20. Part
of the computations were performed on an Linux-Opteron cluster supported by
HBFG-108-291.

\end{document}